*Article*

# Tinnitus, lucid dreaming and awakening. An online survey and theoretical implications.

**Robin Guillard** [1,2*], **Nicolas Dauman** [3,4], **Aurélien Cadix** [2], **Charlotte Glabasnia Linck** [2], **Marco Congedo** [1], **Dirk De Ridder** [5] and **Alain Londero** [6]

[1]  GIPSA-Lab , Univ. Grenoble Alpes, CNRS, Grenoble INP, Grenoble, France.
[2]  Robin Guillard EIRL, Grenoble, France.
[3]  Université de Poitiers, Univ Rennes, Univ Angers, Univ Brest, RPPSY, Poitiers, France.
[4]  Maison des Sciences de l'Homme et de la Société, Université de Poitiers – CNRS, Poitiers, France.
[5]  Department of Surgical Sciences, Section of Neurosurgery, Dunedin School of Medicine, University of Otago, New Zealand
[6]  Université Paris Cité, Institut Pasteur, AP-HP, Hôpital Lariboisière, Service ORL, INSERM, Fondation Pour l'Audition, IHU reConnect, F-75010 Paris, France

Correspondence: robin.guillard@grenoble-inp.fr, 17 boulevard de Picpus, 75012, Paris

**Abstract:** (1) Background: Tinnitus is the perception of phantom sound in the absence of a corresponding external source. Previous studies reported that the presence of tinnitus is notably absent during dreams. This study aimed at replicating previous findings regarding tinnitus-free dreams, while also gaining a deeper understanding of tinnitus manifestations during dreams and after awakening.

(2) Methods: For this observational study, 195 tinnitus patients answered an online survey on the mutual-help community Siopi.

(3) Results: 160 patients could recall their dreams. Among them, 92.5% state they do not hear their tinnitus while dreaming. The rest (7.5%) report higher tinnitus burden, higher stress and more often exhibit objective tinnitus and/or tinnitus related to peripheral auditory pathology and/or drug intake. 13% of the participants frequently experience lucid dreams. Among them, 36% could perceive their tinnitus during lucid dreams, and this was strongly associated with the concomitant perception of external sounds during lucid dreaming. While the majority of patients report perceiving their tinnitus instantly upon awakening, during nocturnal awakenings, 18% declared they could be awakened by their tinnitus and 9.8% mentioned that their tinnitus can temporarily cease.

(4) Conclusions: Our findings confirm the previous findings: tinnitus is rarely perceived during dreams. Remarkably, our study is the first to document the case of tinnitus during lucid dreaming. 64% of these patients gain higher-order consciousness attributes while still experiencing a tinnitus-free state. Our observations suggest that the presence or absence of gating of external auditory information during dreams acts as a tinnitus on-off switch, refining the previously proposed integrative model of auditory phantom perception.

**Keywords:** Tinnitus, dreaming, REM sleep, lucid dreaming, awakening

**Highlights:**

-    Only 7.5% tinnitus patients report hearing tinnitus during their dreams

-    This subgroup reports higher tinnitus burden and higher stress

- 36% tinnitus patients experiencing lucid dreams report hearing tinnitus during them

- During lucid dreams, tinnitus and external sounds perception are strongly associated

- 22% patients both experience increased tinnitus awakening from dreams and from naps

## List of abbreviations

MEMA : Middle-Ear Muscle Activity
PGO waves: Ponto-Geniculo-Occipital waves
REM : Rapid Eye Movement
SWS : Slow Wave Sleep
TDA survey : Tinnitus, Dreams, and Awakening survey
THI : Tinnitus Handicap Inventory
VAS : Visual Analogic Scale
VNS : Visual Numeric Scale

## 1. Introduction

Tinnitus is defined as "the conscious awareness of a tonal or composite noise for which there is no identifiable corresponding external acoustic source" (De Ridder et al., 2021). This chronic symptom is widely prevalent, affecting approximately 14% of the world population (Jarach et al., 2022). Although this symptom is a severe burden for 2% of the population (Jarach et al., 2022), recent studies have shown that tinnitus is most often absent during dreams (Aazh et al., 2021; De Ridder et al., 2014a). Indeed, two studies report that around 95% of the tinnitus population who remember at least some of their dreams do not report hearing their tinnitus during dreams, although tinnitus can constitute a daily burden for these individuals during wakefulness. Two studies (De Ridder et al., 2014a; Hullfish et al., 2019) have proposed that the peculiar phenomenon is consistent with their theoretical models of tinnitus, the Bayesian brain model and predictive coding. These models share many similarities and both posit that tinnitus results from a prediction error in auditory perception. Additionally, both studies argue that the absence of tinnitus during dreams can be attributed to the intact body representations that are typically present in dreams (Mulder et al., 2008), leading to the virtual experience of perfect hearing and, consequently, the absence of tinnitus.

Yet, these articles do not address why a minority of patients report that they sometimes hear their tinnitus in their dreams. In their study, Aazh and colleagues (Aazh et al., 2021) provide initial insights into this issue by examining the symptomatologic features of the subgroup of patients who report auditory experiences of tinnitus in their dreams. Their findings indicate that these patients have a significantly higher tinnitus burden and are marginally younger in age.

Moreover, none of these articles evaluates how tinnitus behaves in the specific setting of lucid dreaming. A lucid dream occurs when one becomes aware of the fact that one is dreaming while continuing to dream (Baird et al., 2021). It is generally considered that lucid dreaming is a dissociated state of consciousness between wake and non-lucid (classic) dreaming (Voss et al., 2009), yet a recent study proposed that it is not a mixture of wakefulness and sleep but occurs in a state of activated Rapid Eye Movement (REM) sleep (Baird et al., 2022).

Lucid dreaming is a state associated with a restauration of high order consciousness attributes, that is to say advanced cognitive functions, such as reflexive consciousness, metacognition, volitional access to memory, and the ability to modulate attention (Baird et al., 2019). Notably, it appears that some lucid dreamers are able to process external sounds and tactile stimulations during their lucid dreams and respond to these stimuli. In one study, the experimenters performed an P300 auditory evoked potential task with an oddball paradigm consisting of sound stimulations of either 1000 Hz or 2000 Hz (Appel and Pipa, 2017). Among three lucid dreamers who performed 10 verified lucid dreams, oddball sounds were detected during eight lucid dreams with a success rate varying between 27% and 100%. When the oddball sound was detected, a P300 wave was detected associated with it, it was absent when not detected by the lucid dreamer. In another study, researchers gathered a collection of results illustrating that two-way communication was possible with lucid dreamers, notably when the task consisted in responding to simple math problems transmitted either through oral modality or tactile stimulations (Konkoly et al., 2021). Two-way communication was not always successful, this collection of reports concludes that on 158 trials, correct answers to the questions asked to the lucid dreamer under polysomnographic control, correct answers were given in only 18,4%.

Such low success rate in tasks that are relatively simple to execute while awake can be attributed to the phenomenon that, during lucid dreaming and sleep in general, external stimuli frequently fail to access consciousness. The classical view concerning gating of sensory information during sleep states that such gating occurs at the level of the thalamus and may be orchestrated by the inhibitory role of thalamic reticular nuclei (McCormick and Bal, 1994). Although there is now substantial empirical evidence that some external stimuli are processed by the sleeping cortex, this model is still supported in the present days by some authors (Coenen, 2022). On the other hand, some authors recently proposed a novel model for the gating of sensory information during sleep (Andrillon and Kouider, 2020). Interestingly, these authors propose that the brain may perform such sensory gating differently depending on whether the sleeper is in slow wave sleep (SWS which comprises N2 and N3 sleep stages) or in REM sleep. For SWS, they rather refer to cortical gating than to thalamic gating, supporting that the information reaches the cortex but is not accessed by the consciousness-enabling network. However, during REM sleep, it is suggested that there is an information filtering process that takes place, where the individual is diverted from external auditory stimuli due to the internally-created perceptions of their dreams.

Such distinction between gating processes depending on the sleep stage matches observations of the functioning of auditory neuron units along the central auditory pathway. Indeed, in unit measurement studies on rodents, spontaneous and evoked unitary firing patterns were found different between wake, SWS and REM sleep in the cochlear nuclei (Pena, 1992), superior olivary complex (Pedemonte et al., 1994), inferior colliculus (Torterolo et al., 2002), Medial Geniculate Body (MGB) (Edeline et al., 2000) and auditory cortex (core, belt and para-belt) (Edeline et al., 2001; Issa, 2008). Generally, what has been observed are reduced spontaneous activity during SWS with higher bursting synchronized activity and evoked activity, whereas spontaneous activity is higher during REM sleep, sometimes even higher than during wake, with a lower signal to noise ratio for the processing of external auditory information from the inferior colliculus to the auditory cortex.

Synchronized bursting activity (Noreña and Eggermont, 2003) and higher spontaneous activity (Kaltenbach and McCaslin, 1996; Munguia et al., 2013) in the central auditory pathway have both been proposed as alternative neural correlates of tinnitus perception. After a night of sleep, some patients have reported experiencing different levels of tinnitus intensity depending on days, with sometimes what they refer to as a "morning roar" as reported in (Probst et al., 2017). This phenomenon may be attributed to the persistence of certain neural correlates of sleep upon awakening, a concept known as sleep inertia (Trotti, 2017; Vallat et al., 2019). The variation in the timing of awakening either in SWS or REM sleep could potentially explain this clinical observation. Indeed, associated specific REM or SWS sleep inertia correlates in the auditory pathway could modulate differently their tinnitus perception. This interaction between the neural activity during sleep and the neural correlates of tinnitus has been previously proposed in another theoretical review (Milinski et al., 2022).

In the present study, we sought to refine our understanding of tinnitus perception with regard to dreaming, notably by investigating the question of lucid dreaming. We also investigated the manifestation of tinnitus during awakening.

## 2. Materials and Methods:

### 2.1. Participants :

The present study consisted in the completion of an online questionnaire within a group of individuals experiencing tinnitus. The Siopi platform was utilized for this purpose, serving as a digital platform accessible through a mobile application. This platform facilitates mutual help among tinnitus sufferers with similar symptoms and enables them to engage in scientific research. Users of Siopi were notified via email about the opportunity to take part in a research study focusing on tinnitus, dreams, and awakening (TDA). The TDA questionnaire only contained questions about this topic, several demographic data presented hereafter such as age, gender and tinnitus duration were part of other optional questionnaires that patients could decide whether or not to additionally complete, which resulted sometimes in missing values for associated demographic data.

In total, 195 Siopi users answered to the survey between July 2023 and May 2024. The answers of one patient were excluded as he/she declared not perceiving any tinnitus. The sample comprised 71 women and 105 men (18 missing values), with a mean age of 50.98 ± 13.49 years (18 missing values), and a mean time since tinnitus onset of 7.98 ± 9.17 years (48 missing values). Demographic and symptomatologic characteristics of the sample are presented in tables 1 and 2. Initial iteration of the TDA questionnaire did not contain questions about perception of tinnitus specifically during lucid dreaming, only the question about doing or not lucid dreams. After receiving spontaneous valuable testimonies on patient experiences of tinnitus during lucid dreaming, two additional questions were sent to participants who responded they did frequently lucid dreams about their perception of tinnitus during lucid dreams and about the perception of external sounds during lucid dreams.

The question about tinnitus variation pattern was included in both the TDA survey and optional questionnaires. As it corresponds to a general tinnitus symptom characteristic, it was included in the Table 2.

These demographic characteristics presented contain missing data as the questions items they refer to were not directly included in the TDA survey but were part of the questionnaires patients could voluntarily answer to on Siopi. No respondents spontaneously reported difficulties in filling out the questionnaires.

### 2.2. Ethics :

All Siopi users are informed and asked for their consent to accept the general conditions of use and data management policy of Siopi which state that all health-related data collected on Siopi, such as responses to the present survey, can be anonymized and then exported for academic research purposes. Responses were collected in a database hosted in France by an approved health-data center, and are exported separately from personal data so as to ensure strict anonymity. The email campaign advertising the survey clearly mentioned that the purpose of such survey was to perform a scientific study leading to a publication.

### 2.3. Statistical analyses:

Siopi, in charge of data processing, exported the anonymized responses of the responders of the survey for analysis from the secure database in comma separated value (CSV). This CSV database was then analyzed using Anaconda Python 3.7 software and Jupyter Notebook module using Python, pandas library (McKinney, n.d.) and Scipy library (Virtanen et al., 2020).

To identify the symptomatologic characteristics of the subgroup of patients declaring they remembered some of their dreams and perceived tinnitus during their dreams, we used a Mann-Whitney test to compare the characteristics of this subgroup with the group of patients that remembered some of their dreams and that did not perceive it. This analysis was conducted on each of the features of the dataset, and also calculated an associated Hedge g effect size (Hedges, 1981). The same methodology was used to perform a supplementary analysis to test if there was an association between increased tinnitus perception waking from a dream and nap-induced tinnitus modulations. For these analyses, due to the rarity of patients declaring perceiving tinnitus during their dreams and the abundance of features to be tested, we decided not to perform statistical corrections.

To analyze whether there was a difference between the characteristics of tinnitus during awakenings between during the night (nocturnal awakening) and in the morning, we used a paired Wilcoxon test on the answers given on the 2 questions on nocturnal awakenings and morning awakenings. Benjamini Hochberg false discovery rate statistical corrections was applied to correct for multiple testing on this set of statistical tests (Benjamini and Hochberg, 1995).

### 3. Results :

The answers to the TDA survey are presented in Table 3. Cross-observations on answers to questions about perceiving tinnitus and/or external sounds during lucid dreams are presented in the Table 4. In this cross-observation table, only patients who could report on both questions were selected (patients who reported "not concerned/not being able to report" to either question or both were excluded).

It is important to notice through tables 3 and 4 that the proportion of patients declaring perceiving their tinnitus during lucid dreams (8 / 22, 36.3%) is higher than the proportion of patients perceiving their tinnitus during normal dreams ( 12 / 160, 7.5 %). Interestingly, all but one of the patients declaring perceiving tinnitus during lucid dreams also declared being able to hear external sounds during lucid dreams. Conversely, all but one of the patients who declared not hearing tinnitus during their lucid dreams (14 / 22, 59.1%) also declared not being able to hear external sounds during their lucid dreams. As a consequence, for 20 / 22, 90.9% of the sample, perceiving tinnitus was associated with perceiving external sounds during lucid dreams.

The significant symptomatologic characteristics of the subgroup of patients hearing tinnitus during their dreams is summarized in Table 5. Results are ordered in terms of effect sizes. A positive effect size refers to a characteristic more present in the group of patients hearing their tinnitus during their dreams while a negative effect size refers to a characteristic more present in the rest of the sample. Sample sizes for each test are specified as for each of the symptomatologic characteristic evaluated the crossing of missing data can be different. In table 5, only tinnitus and comorbidities related features are reported (no lifestyle habits, nor treatment decision nor TDA questionnaire items), in case of redundancy between features of different standardized questionnaires exhibiting significative results, the feature presented is the one with the least missing values.

As illustrated by table 5, the subgroup of patients perceiving tinnitus during their dreams is convergently characterized by: exhibiting significantly more often objective forms of tinnitus or peripheral hearing pathologies; having a significantly higher likelihood of presenting with important hearing impairment and/or with initial tinnitus emergence associated with drug intake (aspirin, quinin, antibiotic). They also present significantly higher degree of stress/anxiety, tinnitus impact on quality of life, depressive symptoms and sleep disturbances.

The differences between tinnitus perception upon nocturnal awakening and upon morning awakening is illustrated in Figure 1. Significantly more responders declared tinnitus appeared gradually following morning awakening compared to nocturnal awakening (p < 0.05) and significantly more responders declared not perceiving their tinnitus during nocturnal awakening compared to morning awakening (p < 0.05).

A supplementary analysis was conducted, mirroring the previous investigation aimed at characterizing the symptoms of tinnitus in patients who experience tinnitus during their dreams. We investigated patients who reported increased tinnitus loudness following dreams and tested how it was related to nap-induced tinnitus modulation. This supplementary analysis indicated that this subgroup of patients exhibited a significantly greater incidence of increased tinnitus after naps until their next sleep episode (Mann-Whitney test, p < 0.001, g = 0.6).

## 4. Discussion

This study sought to enhance comprehension of the correlation between tinnitus and dreams, as well as between tinnitus and awakening. It effectively replicated previous findings from (Aazh et al., 2021; De Ridder et al., 2014a) regarding the percentage of tinnitus sufferers who reported not experiencing tinnitus during dreams. It provided a more comprehensive insight into the symptomatology of tinnitus patients experiencing tinnitus during their dreams, by adding the experience of tinnitus patients during lucid dreams, and by offering a more detailed understanding of how tinnitus manifests when awakening from a nap, a dream, in the middle of the night, or in the morning.

*Tinnitus and dreams*

In the present study, 92,5% of the 160 responders who reported remembering at least some of their dreams stated that they did not perceive tinnitus during their dreams, which is in accordance with past studies (Aazh et al., 2021; De Ridder et al., 2014a). Additionally, 83.1% of 160 patients reported that the concept of tinnitus was not topic in their dreams despite the fact our sample consisted of tinnitus patients bothered by their tinnitus (THI score 58.95 +/- 23.1).

The presence or absence of tinnitus in the recall of patients' dreams rises questions which are at the crossroads of divergent frameworks in the dream literature. There are two main hypotheses in the field. On the one hand, the 'discontinuity' hypothesis argues that the dreaming consciousness is *isolated* from the awakening experience. On the other hand, the 'continuity' hypothesis claims that one *incorporates* (part of) the dreamer's awakening experiences into their dreams. For review, see (Windt, 2018).

The 'discontinuity' hypothesis is supported by the fact that the majority of our sample report no tinnitus in their dreams recall. In fact, the absence of tinnitus in dream remarkably fits with the singularities of dreaming consciousness, as contrasted with wakening consciousness (D'Agostino, 2013; Hobson, 2009; Windt, 2010). Although cognition (Graveline and Wamsley, 2015; Kahan, 1996) was claimed to contribute to the dream content, a distinctive feature of dreaming consciousness is the ''*strong tendency for a single train of related thoughts and images to persist over extended periods without disruption or competition from other simultaneous thoughts and images*'' (Rechtschaffen, 1978) (p. 97). In other words, being fully immersed in any 'events' passing before awareness, the dreaming consciousness is most often unable to critically reflect about it (*i.e.*, this is only a dream, not the reality) or draw attention to thoughts apart from what is currently happening (*i.e.*, more relevant thoughts). In fact, the dreaming consciousness is confined in the here-and-now of the dream content (D'Agostino, 2013; Windt, 2010). While such immersion is only partial in daydreaming (Windt, 2018), isolation of dreaming consciousness from most sensory inputs makes it rather inaccessible to the dreamer's reflectiveness (Rechtschaffen, 1978).

Interestingly, the dream extended periods without interruption echoes the *flow* state of wakeful consciousness (Csikszentmihályi, 1990) that was identified in patients' qualitative reports of tinnitus dissipation, when they are engaged with rewarding activities (Dauman and Dauman, 2021). Consistent with the limited experience of time in dream (*i.e.*, a sense of duration only, without time perspective), the flow state is characterized by the individual losing track of time while being fully-absorbed in the activity at stake (see (Dauman, 2023) for a theoretical framework on time perception in tinnitus patients). The common absence of tinnitus in dreaming consciousness (apart from nightmares, see below) could be underpinned by full immersion in the dream images and thoughts, which is harder to achieve in wakeful activities because of tinnitus-induced frustration (*i.e.*, self-monitoring and ruminations, see (Dauman et al., 2017)).

The minority of patients who report hearing their tinnitus during their dreams is in alignment with the predictions of the 'continuity' hypothesis. Indeed, it would make sense that the salience of tinnitus in the life of suffering patients would make it to be incorporated into their dreams, as reflecting a major personal concern (Domhoff, 2015), associated with high emotional intensity (Schredl, 2006, 2003). It has additionally been shown that higher levels of anxiety (Rimsh, 2022) and feelings of loneliness (Paul, 2021) enhance the rate of incorporation of waking life experiences into dreams (*i.e.*, nightmares associated with a sense of helplessness in the dreamers).

The symptomatologic characterization of tinnitus patients reporting tinnitus during their dreams supports these traits. In a former study, the subgroup of patients reporting hearing tinnitus during their dreams has been associated with experiencing significantly higher impact of tinnitus on quality of life and appeared marginally younger patients (Aazh et al., 2021). In the present study, we confirm that this subgroup presents significantly higher tinnitus impact on the quality of life (measured by THI and a VAS scale on tinnitus annoyance). That tinnitus tends to be incorporated in patients' recalls who report higher burden and level of stress aligns with the putative role of somatosensory aversiveness in shaping dreaming consciousness (*i.e.*, 'hearing' or 'thinking about' tinnitus in dreams). We also find new dimensions to characterize this subgroup as they exhibit significantly more often objective forms of tinnitus or associated with, peripheral hearing pathologies; significantly more important hearing impairment. Additionally, they report significantly more often initial emergence associated with drug intake (aspirin, quinin, antibiotic). They also present significantly higher degree of stress/anxiety, depressive symptoms and sleep disturbances. To a lesser extent, we also found significantly higher reported vertigo ear fullness and headaches, possibly associated with reported higher rate of peripheral hearing pathologies.

In perspective, the fact that tinnitus can be present or absent during dreams is similar to what has been recently reported for limb amputation in the literature (Bekrater-Bodmann, 2015), which challenges the initial hypothesis of an innate body representation seemingly 'insensitive' to such dramatic alterations (Mulder et al., 2008). In this large study (Bekrater-Bodmann, 2015), almost half of the sample (N=3234) was either unable to remember their body representation in their dream (22%) or dreamt about themselves as intact (24%). In only 3% of the reports was amputation always incorporated into dreams and, for 1/3 of the amputees, the body image could be either intact or impaired. Thereby, the authors suggest, in addition to physical alteration (*i.e.*, phantom limb), pain-related suffering (*i.e.*, aversiveness) would underly the incorporation of amputation in dreams.

Finally, it is noteworthy to mention that in the present study, participants distinguished two aspects in the incorporation of tinnitus in their dreams. While only 7.5% of them reported being able to hear tinnitus during their dreams, more than the double of this proportion (16.9%) reported that they made dreams *about* their tinnitus. Interestingly, such a distinction echoes patients' worries about their awareness of "having tinnitus" (as a feature that defines self-perception) even though they do not hear it, in fact, while immersed in enriched sound environments (Dauman et al., 2017). Those patients also report being prone to monitoring the presence of tinnitus when they do not hear it (with the hope tinnitus has 'disappeared'). Thus, thinking about tinnitus would be partly dissociated from its actual presence as auditory percept, both in wakefulness and in dreaming consciousness. Further investigations of the incorporation of tinnitus into dreams should carefully phrase questions and check for participants' understanding of these qualifications.

*Tinnitus and nightmares*

Nightmares are an exception to the attention-captured pattern in dreaming consciousness, for they rather exacerbate the salience of tinnitus in patients' recalls. Dream literature suggests an argument which would both fit with the absence of tinnitus in most dream recalls and its incorporation into vivid nightmares. This argument concerns the *reduced form* of self in dreaming consciousness, that is approximately assumed to be equivalent to self-perception in wakefulness (Windt, 2010). However, if visual information is most often present in dreams, many aspects of bodily experience are also systematically underrepresented, such as vestibular, thermal or touch sensations. As contrasted with our full-fledged body experience, the body representation in dreams is seldom detailed (*i.e.*, with distinct body parts), nor fully integrated as a whole experience (Windt, 2010). Thereby, a more suitable definition of the dream self would be that of a ''*purely spatiotemporal first-person perspective*'' (Windt, 2010) (p. 313), which accounts for the sense of self-location and duration in relation to the dream content. Like intact body representation in amputees' dreams (Bekrater-Bodmann, 2015) or able-bodied experiences in congenital paraplegics (Koppehele-Gossel et al., 2016), the absence of tinnitus *as a bodily feature of the dream self* would result from the paucity of self-representation (with no distinct parts or impairments) and information that should be integrated by the auditory perception during sleep (Hullfish et al., 2019).

Nightmares are typically characterized by the violation of the dream *self-integrity*. In most instances, the dreamer is the target of an ongoing physical aggression or a persistent threat, such as being chased or confronted with an evil presence (Mathes et al., 2019; McNamara et al., 2015; Paul, 2021). Thus, nightmares occur when perceived distance with potential threat is breached, which denotes a minimal sense of bodily *boundaries* that must be readily secured. Contrary to common dreams (including bad dreams about interpersonal conflicts (Paul, 2021)), nightmares result in the disruption of dreaming consciousness by abrupt awakening.

In the present study, the symptom most closely linked to experiencing tinnitus during dreams was stress/anxiety. Indeed, this is corroborated by the observed answers to our questionnaire (Table 3) which exhibit that dreams associated with tinnitus are more often nightmares than positive dreams (hence more stressful). The survey results also point out that a subset of participants mentioned that waking up from these dreams appears to result in increased tinnitus loudness.

The combination of these observations indicates that the occurrence of tinnitus during dreams, potentially leading to awakening due to tinnitus, and the increased severity of tinnitus upon waking from a dream may be associated with a minimal sense of self boundaries (*i.e.*, self-integrity) that is induced by threat-related stress during nightmares. This heightened stress has been proposed to involve the reactivation of adrenergic neurotransmission during dreaming (involving serotonin, noradrenaline, dopamine), despite the fact that these systems should be deactivated during REM sleep (Brezner, 2011; Pagel and Helfter, 2003).

A simplified explanation for the awakening from a tinnitus 'nightmare' could be proposed by the analogy with PTSD. In a subgroup of patients in whom the development of tinnitus is perceived as a traumatic experience, it can be expected that analogous to awakenings due to trauma-related nightmares, tinnitus-related awakenings may ensue. If this hypothesis is correct, in this subgroup of patients, a treatment analogous to PTSD may prove beneficial.

*Tinnitus and lucid dreams*

Lucid dreams in tinnitus patients provide a unique set of observations that shall take place between tinnitus percept in wakefulness and, in most instances, its absence in common dreams. The consciousness that accompanies lucid dreams has been labeled a 'hybrid state' between waking and dreaming (Voss et al., 2009), although this term may be misleading for it suggests too close a proximity with sensory-based self-perception (Baird et al., 2022). Lucid dreamers remain isolated from sensory inputs, but they recover *access* to metacognitive skills that are typical of wakefulness (and are broadly bracketed in common dreams, see (Hobson, 2009; Rechtschaffen, 1978)). These metacognitive skills include volition, self-reflection and access to waking memories (Stumbrys et al., 2014). As contrasted with deluded sense of wakefulness in common dreams, lucid dreamers are aware that their experience is hallucinatory (D'Agostino, 2013). In addition, they have a sense of agency upon their dreams, for they can influence the dream content to a certain extent (Stumbrys et al., 2014) although enjoyment also derives from simply observing the dream in detail (Mallett et al., 2022). However, lucidity in dreaming consciousness does not always result in successful achievement of actions that would be impossible in waking life (*e.g.*, flying). On the contrary, less than 50% of intentions would be successfully accomplished in lucid dreaming, with awakening and hindrances in the dream being common reasons for lack of success (*e.g.*, dream characters do not allow the dreamer's intention) (Stumbrys et al., 2014). Thus, lucid dreamers encounter limits in the enjoyment they derive from controlling the plot of their dream, which comes with the risk of failure or frustration.

Two important findings seem to emerge from the observations made from the survey results on this topic.

First, we observed that to hear one's tinnitus during a lucid dream (8 / 22, 36.3%) appeared to be more frequent than hearing tinnitus during common dreams (12 / 160, 7.5 %). This suggests that the dynamics surrounding the dreamer's sense of agency would play a role in the chances of reemergence of tinnitus. In particular, the more the dreamer would encounter hindrances (*i.e.*, frustration) in his/her attempts to influence the dream plot, the more likely he/she would experience tinnitus reemerging in self-awareness. To explore this new area in tinnitus, additional exploratory interviews were conducted (ND) with some participants who were identified as lucid dreamers. While lucid dreams were in most instances tinnitus-free, participants also indicate perceiving a reemergence of tinnitus in parallel with an intensified intention to achieve their goal. For instance, one participant reported increased loudness of tinnitus in lucid dreaming correlating with her desire to fly higher and higher. She eventually stopped flying because of the intrusiveness of tinnitus in her dream. Another participant reported that tinnitus reemerged as a percept in lucid dreaming following a nightmare in which he was pursued by undefined people who wanted to hold him back. These preliminary observations suggest that heightened self-awareness, because of negative emotions such as fear or frustration, would increase the risk of hearing tinnitus in lucid dreams. A qualitative analysis of these findings is presented in the Supplementary material.

The results presented in Table 5 highlight a second finding. Indeed, it appears that hearing tinnitus during a lucid dream is strongly associated with being able to perceive external sounds during one's lucid dream. In other terms, this suggests that the presence or absence of gating of external auditory information during dreams could testify whether tinnitus is heard or not. Yet, this finding has to be nuanced considering the patient interviews presented in supplementary material. In some of these patients reports, tinnitus could emerge during lucid dreams, and potentially even during some nightmares, when the dreamer is confronted with the encounter of hindrances in the dream plot.

However, it is noteworthy that only 36,3% of patients are capable of perceiving external sounds during their lucid dreams, which is not unexpected. Indeed, this is comparable to the proportions observed in previous studies, one exploring P300 responses during lucid dreams (Appel and Pipa, 2017) and the other exploring two-ways interaction with the lucid dreamer (Konkoly et al., 2021).

The association between the perception of both external sounds and tinnitus during lucid dreams may lead to the formulation of additional hypotheses.

According to the informational gating theory (Andrillon and Kouider, 2020), during dreaming, external sensory information is prevented from reaching consciousness due to the dominance of competing internally-generated sensory information. This theory suggests that two competing streams of auditory information coexist during dreams. Our findings strongly indicate that when the lucid dreamer exclusively focuses on the internally-generated/oneiric auditory stream of information, tinnitus is not experienced. Conversely, when the lucid dreamer shifts attention to external sounds or when the gating of external information fails, leading to exposure to external sounds, then tinnitus reemerges. Figure 2 summarizes the heterogeneity of tinnitus experience in dreaming and waking consciousness according to levels of self-awareness, providing a first account of the observations discussed in the present study.

The finding that tinnitus is perceived during lucid dreams when also external sounds are perceived is in keeping with the Bayesian brain model of tinnitus. The predictive Bayesian brain model states that during the awake state the brain constantly makes predictions about the environment(De Ridder, Vanneste et al. 2014). Tinnitus is hypothesized to be the result of a prediction error due to deafferentation, and missing input is filled in by the brain (De Ridder, Joos et al. 2014, De Ridder, Friston et al. 2023). The heuristic explanation then is that in the dream state there is no interaction with the environment and therefore no updating of the prediction error, resulting in the absence of tinnitus(De Ridder, Joos et al. 2014). Thus, lucid dreams without external auditory perception are akin to normal dreams, in which no tinnitus is heard, and lucid dreams with external auditory perception are akin to wakefulness, in which tinnitus is heard (Figure 3)

While the Bayesian brain theory and the predictive coding theory applied to tinnitus bring compelling arguments on why internally-generated sounds in dreams do not convey tinnitus perception (De Ridder et al., 2014a; Hullfish et al., 2019), these models do not provide insight on the pathway used by such internally-generated sounds during our dreams. In other terms, one can ask oneself: where do the sounds of our dreams come from?

*Where do the sounds of our dreams come from?*

We led an investigation of past literature on this topic as this question was raised from our observations on tinnitus lucid dreamers. The results of this bibliographic exploration are summarized hereafter.

It has been suggested (Andrillon and Kouider, 2020) that during REM sleep informational gating occurs, in which the brain would be "distracted" by the emission of Ponto-Geniculo-Occipital (PGO) waves by the reticular nuclei and would thus not be able to properly process external sounds. They suggested that such gating should operate at the cortical level. Yet, by exploring the scientific literature on the unitary firing properties of the auditory nuclei during REM sleep, another slightly different alternative hypothesis can be posited. Indeed, such studies have shown that spontaneous unitary firing properties in the inferior colliculus, medial geniculate body and auditory cortex (A1, belt and parabelt) are globally increased during REM sleep (Edeline et al., 2000; Issa, 2008; Torterolo et al., 2002). At the same time, unitary evoked activity by clicks remains unchanged in amplitude (contrary to Non-REM sleep). This results in a decreased signal to noise ratio between evoked responses and spontaneous activity, which has been described as significant in at least two studies (Edeline et al., 2000; Torterolo et al., 2002). These studies together lead us to propose another hypothesis: there seems to be a form of informational gating of auditory information during REM sleep, but it may not only take place at the cortical level. Rather, it could start in sub-cortical nuclei, at least down until the inferior colliculus and maybe even down to the cochlear nuclei. One may wonder what would be the cause of the enhancement of such spontaneous activity in the central auditory pathway during REM sleep.

A good potential candidate would indeed be auditory PGO-like waves, as presented in (Fazen, 1987). Such PG-auditory waves would be associated with middle-ear muscle activity (MEMA) which would appear as the phasic auditory equivalent of rapid-eye movements during REM sleep. Interestingly, it would match the theory proposed by Hobson and Friston (Hobson and Friston, 2012) on the role of rapid-eye movements and PGO during REM sleep. In fact, they posit, according to the Bayesian brain framework, that during REM sleep the brain optimizes our prior belief model. Physiologically, it would rely on the deactivation of the supra-granular layer of the sensory cortices by an interruption of aminergic excitation. This would lead to the interruption of the computation of prediction error at the level of the sensory cortices. Yet, they mention that on the contrary proprioceptive processing of the muscles associated with sensory acquisition, such as the orbital muscles of the eyes, remain active and even enhanced during REM sleep and would actually substitute real sensory information processing during REM sleep. Such muscle movements would be activated by PGO waves. Proprioceptive information would hence be normally processed (and even enhanced) and be interpreted at the level of the sensory cortices as sensory information that would take meaning in the context of the dreams and contribute to prior belief model optimization. They describe this view as a revisited interpretation of PGO waves and thoroughly describe it for the visual modality. The authors did not propose that a similar functioning could take place for the auditory modality as they maybe were not aware of specific REM sleep MEMA activity. Indeed, MEMA activity could truly be the equivalent of rapid-eye movements for the visual modality during REM sleep. They could be elicited by PGO-like activity generated by the reticular nuclei and then propagate within the central auditory pathway, resulting in the observed enhanced spontaneous activity. Proprioceptive information from MEMA could be enhanced such as what happens in the visual modality. This would result in a decrease of signal to noise ratio of real sound processing, as was reported in the inferior colliculus (Torterolo et al., 2002), medial geniculate body and auditory cortex (Edeline et al., 2001, 2000). This proprioceptive signal would then be interpreted as sensory auditory information at the level of the auditory cortex in the context of the dream. Such proprioceptive information would be somehow amorphous and could account for intact body representation during REM sleep as it would not reflect any tonotopically-restricted deafferentation.

This would result in the absence of tinnitus perception associated with such stream of internally-generated auditory information. We could add that in such a view, and considering the predictive coding model applied to tinnitus, the accumulation of REM sleep may be beneficial to tinnitus patients and may make tinnitus remission more likely. Indeed, during REM sleep, the prediction model trains itself to fit an intact hearing representation instance, devoid of tinnitus perception. This could partly explain why several studies found that REM sleep decrease is associated with tinnitus presence (Attanasio et al., 2013; Guillard et al., 2023; Teixeira et al., 2018).

Anecdotally, we could mention that if an individual presented a lesion only at the level of the middle-ear muscles and yet have an intact auditory system (cochlea and auditory nerve), one could presume that inversely, REM sleep may have a negative effect on hearing and could potentially lead to tinnitus emergence by using deteriorated proprioceptive information as base for auditory prediction model fitting. Such a set of hypotheses deserves further clinical investigations.

### Lucid dreams and the integrative model of auditory phantom perception

The integrative model of auditory phantom perception (De Ridder et al., 2014b) suggests that "*the tinnitus on/off switch should be located in one or more areas that differ between waking and REM state*" and used this criterion to define the "*tinnitus core network*", consisting of the auditory cortex, inferior parietal region, the parahippocampus and the ventrolateral prefrontal cortex. The observation that tinnitus perception is strongly associated with external sound perception during lucid dreams comforts and precises such a vision on location of the tinnitus on/off switch location. We hence propose Figure 4 as a revised version of their former Figure 3 following our observations. According to our adjusted view of the model, tinnitus on/off switch would be located in the difference of brain characteristics between the state of lucid dreaming with or without external sound perception. It remains an open question whether the lighting this switch is associated with a failure of the auditory gating system during lucid dreaming sleep, with a progression in the degree of lucidity one may acquire within the dream, or both.

### Tinnitus and awakening

As presented in the introduction, through sleep inertia, awakening from SWS or REM sleep could lead to the transient maintenance of respectively synchronized bursting activity or higher spontaneous activity in the central auditory pathway. This could lead to modulated tinnitus perception upon awakening. One could posit that REM sleep inertia has higher probability of occurring if one is interrupted in the middle of a dream, and even more in the case of a vivid/intense dream. On the other hand, previous studies support that patients report sometimes intense sleep inertia following an afternoon nap of 45 min or less (Tassi and Muzet, 2000; Vallat et al., 2019), in which except for narcoleptic patients REM sleep is normally absent (Lewis, 1969). Such reasons led the authors to include in the questionnaire two questions regarding tinnitus perception upon awakening from a dream and from a nap. The responses to these inquiries yielded similar outcomes: while most participants did not notice any changes in their tinnitus after waking up from a dream or a nap, among those who did report a difference, a higher number of individuals reported an increase rather than a decrease in tinnitus perception.

The observation that patients reporting increased tinnitus loudness following dreams exhibited a significantly greater incidence of increased tinnitus after naps until their next sleep episode suggests that they may represent one unique subgroup of patients. These results tend to indicate that both synchronized bursting activity and elevated spontaneous activity within the central auditory pathway may contribute to an enhanced perception of tinnitus in at least a subset of these patients.

Conversely, regarding the manifestation of tinnitus upon waking during the night and in the morning, the findings of the current survey indicate that most patients experience tinnitus immediately upon awakening in both scenarios. Additionally, the data reveal that, while it occurs in a minority of instances, patients are less likely to report being awakened by their tinnitus compared to those who perceive it after a delay. This delayed manifestation of tinnitus appears to be variable (progressive or brutal, linked to attentional focus, tinnitus may even remain absent in some cases (Vanneste, Byczynski et al. 2024)).

A theoretical review (Milinski et al., 2022) proposed that intense SWS in the beginning of the night would alter tinnitus signal in the brain, resulting in tinnitus signal being suppressed. Although the finding that tinnitus seems significantly more often absent during nocturnal awakenings than during morning awakenings (*i.e.*, shorter after sleep beginning) tend to support this theory, it can nevertheless be questioned by the observation that an important part of the tinnitus population seems to report increases of tinnitus after a nap, that also can contain SWS. These contradictory observations appear hard to conciliate and a puzzling challenge.

### Limits

The main limitation of the present research is that the conclusions are based on retrospective observations of patients using a non-exhaustive questionnaire rather than prospective or objective measurements. Notably, a portion of the study's findings relies on reports from tinnitus patients regarding their dream content, whether lucid or not. Previous studies on sleep and dreams have highlighted the limitations of such post hoc dream observations, including the potential censorship of embarrassing content by the dreamer (Schwartz and Maquet, 2002).

It should also be mentioned that the present study exhibited a limit in the affirmation that in the majority of cases tinnitus would be absent during dreams. Indeed, as 83% of our sample declared they do not dream about their tinnitus and hence would not ask themselves whether or not they perceive tinnitus during dreams, it remains an important and open question if tinnitus signal is absent during dreams or if the absence of its perception is related to the absence of any attempt to measure it.

*Suggestions for future research*

To palliate with the retrospective nature of the reports about tinnitus perception during dreams, future research should aim at organizing a prospective protocol including tinnitus patients who happen to be lucid dreamers. In such a protocol, lucid dreamers would be equipped with a polysomnographic measurement material and tasked with signaling first that they gained lucidity within a dream and then signal to the experimenter whether or not they perceive tinnitus during their lucid dreams.

In future studies on tinnitus and awakening, one may investigate which type of dreams results in higher intensity of tinnitus upon awakening: is it specifically vivid/stressful dreams that trigger such reported increases in tinnitus intensity ? Or is it a subgroup of patients who perceive their tinnitus as a traumatic event, as suggested above? Prospective physiological measures could also aim at identifying if such increase is associated with elevated stress or REM sleep inertia.

Moreover, questions on tinnitus intensity upon nocturnal awakenings should be added to better test the hypotheses proposed by Milinski and colleagues (Milinski et al., 2022). Alternatively, a prospective protocol could try to systematically assess tinnitus intensity before sleep onset, during nocturnal awakenings (if one occurs) and then finally upon awakening to test if indeed awakening earlier in the night leads to lower tinnitus. Indeed, although the TrackYourTinnitus study (Probst et al., 2017) apparently concluded on the opposite that tinnitus intensity appears higher between midnight and 8:00 am, patients could freely decide when to report tinnitus, which may have led to a survival bias scenario in which patients would only report their tinnitus in the middle of the night if it was intense and bothering, potentially refraining them from sleep. Indeed, the distribution of reports depending on the hours in this study clearly illustrates a drastic deficit in the number of reports between midnight and 8:00 am compared to the amounts of reports during the day.

Another potential influence could be the depth of the sleep a tinnitus patient wakes up from. It can be hypothesized that when awakening from N3 or SWS that the sympathetic system activates more than when awakening from REM or N1 sleep, thus being more related to parasympathetic predominance. As stress, and sympathetic nervous system activation have been linked to tinnitus worsening(Fagelson 2007, De Ridder, Langguth et al. 2021, Patil, Alrashid et al. 2023) this may need to be investigated as well.

## 5. Conclusions

The current investigation sought to enhance the understanding of the connections between tinnitus and dreams, as well as between tinnitus and the process of awakening. The findings corroborated and extended previous observations, revealing that over 90% of individuals with tinnitus do not perceive their condition during dreaming. Among the subset of patients who do experience tinnitus in their dreams, there was a notable correlation with a greater impact of tinnitus on their quality of life, increased levels of stress, depressive symptoms, and disturbances in sleep. It seems also that they present more often objective tinnitus and/or associated with peripheral auditory pathologies and/or initiated by drug intake.

Additionally, this study was pioneering in exploring the experience of tinnitus patients regarding their condition during lucid dreaming. The results indicated that tinnitus is more frequently perceived during lucid dreams, with a prevalence of 30% compared to 7.5% in non-lucid dreams. A strong association was also identified between the perception of tinnitus and the awareness of external sounds during lucid dreaming episodes. This is in keeping with Bayesian explanations for tinnitus.

Finally, the research provided a more nuanced understanding of how tinnitus presents itself to patients upon awakening. A specific subgroup of patients reported an exacerbation of tinnitus when waking from both dreams and naps. Moreover, while tinnitus typically manifests immediately upon awakening, patients indicated that it can also emerge in various temporal patterns following awakening.

**Supplementary Materials:** The supplementary material "Qualitative analysis of interviews with lucid dreamer tinnitus patients" is attached to the present manuscript.

**Author Contributions:** Conceptualization, R.G. and N.D.; methodology, R.G., D. D R., N.D. and A.L.; software, R.G. and A.C.; validation, R.G., M.C., N.D., D. D. R. and A.L.; formal analysis, R.G. A.C. and C.G.-L.; investigation, R.G., N.D. and A.L.; data curation, R.G., C.G-L. and A.C.; writing—original draft preparation, R.G. and N.D.; writing—review and editing, R.G., M.C, D. D R, N.D. and A.L.; supervision, R.G., M.C., and A.L.; project administration, R.G.; funding acquisition, R.G. All authors have read and agreed to the published version of the manuscript.

**Funding:** This research was funded by Felicia and Jean-Jacques Lopez-Loreta Fondation grant to R.G..

**Institutional Review Board Statement:** The study was based on an anonymized database extracted from an online survey. Responders gave their informed consent that the anonymized data collected through this survey questionnaire could be used for academic research purposes by accepting Siopi terms of use and data policy. As a consequence, the approval of an ethical committee was not required to perform this study. Data collection and storage complied with European and French guidelines for management of health data (RGPD and storage on French HDR servers).

**Informed Consent Statement:** Informed consent was obtained from all subjects involved in the study.

**Acknowledgements:** The authors would like to thank Mr Louis Korczowski for his role as historic founder of Siopi. Similarly, the authors would like to warmly thank France-Acouphènes and the Siopi mobile app team for their help to encourage patients to install the app and answer the survey.

**Financial disclosure:** Robin Guillard declares that he is shareholder and president in Siopi SAS, and has a professional activity as independent as Robin Guillard EIRL.

The other authors did not declare financial interests.

**Non-financial disclosure:** None.

**Data availability statement:** The data underlying this article can be shared on reasonable request to the corresponding author.

**Declaration of generative AI and AI-assisted technologies in the writing process:** During the preparation of this work the authors used https://ahrefs.com/writing-tools/sentence-rewriter in order to only improve the language and the readability of the present manuscript, with caution and manual supervision. After using this tool/service, the authors reviewed and edited the content as needed and take full responsibility for the content of the publication.

## References

References must be numbered in order of appearance in the text (including citations in tables and legends) and listed individually at the end of the manuscript. We recommend preparing the references with a bibliography software package, such as EndNote, ReferenceManager or Zotero to avoid typing mistakes and duplicated references. Include the digital object identifier (DOI) for all references where available.

Citations and references in the Supplementary Materials are permitted provided that they also appear in the reference list here.

## Table 1, Sample characteristics for quantitative features (N =194)

| | Mean | Std | Min | Max | Missing values |
|---|---|---|---|---|---|
| Age | 50.98 | 13.49 | 18 | 79 | 18 |
| VNS on tinnitus | 6.75 | 2.08 | 1 | 10 | 0 |
| VNS on ear cracking | 2.63 | 3.18 | 0 | 10 | 0 |
| VNS on plugged ears | 3.18 | 3.1 | 0 | 10 | 0 |
| VNS on headaches | 3.27 | 3.07 | 0 | 10 | 0 |
| VNS on sleep disturbance | 5.64 | 2.97 | 0 | 10 | 0 |
| VNS on vertigo | 2.23 | 2.9 | 0 | 10 | 0 |
| VNS on jaw pain | 2.56 | 3.03 | 0 | 10 | 0 |
| VNS on cervical pain | 3.99 | 3.22 | 0 | 10 | 0 |
| VNS on anxiety level | 6.1 | 2.57 | 0 | 10 | 0 |
| VNS on depressive mood | 4.43 | 2.71 | 0 | 10 | 0 |
| THI score | 58.97 | 23.1 | 8 | 100 | 40 |
| ISI score | 13.24 | 6.29 | 0 | 25 | 25 |
| Height (cm) | 172.18 | 12.29 | 150 | 193 | 48 |
| Weight (kg) | 75.91 | 18.59 | 49 | 210 | 48 |
| Duration tinnitus (in years) | 7.98 | 9.17 | 7 | 48 | 48 |
| *VNS : Visual Numeric Scale, THI : Tinnitus Handicap Inventory, ISI : Insomnia Severity Index* | | | | | |

*Table 1 – Sample characteristics for quantitative features, N=194, Abbreviations : VNS : Visual Analog Scale, THI : Tinnitus Handicap Inventory, ISI : Insomnia Severity Index, Std : standard deviation.*

## Table 2, Sample symptomatologic characterisation (N = 194)

| General chracteristics | Percentage | Tinnitus attributes | Percentage | Tinnitus variation | Percentage |
|---|---|---|---|---|---|
| **Gender** | Ms : 18/194 | **Tinnitus perception** | Ms : 48/194 | **Modulation increase *** | Ms : 11/194 |
| Man | 59.7 % | Intermittent | 2.7 % | Very quiet environment | 55.2 % |
| Woman | 40.3 % | Constant | 97.3 % | Low intensity sounds | 8.2 % |
| **Hearing status** | Ms : 48/194 | **Tinnitus laterality** | Ms : 41/194 | High intensity sounds | 32.2 % |
| Cophosis | 2.7 % | Only left | 18.3 % | Head movements | 13.1 % |
| Severe hearing loss | 17.8 % | More left than right | 15.7 % | Clenching the teeth or moving the jaw | 24.6 % |
| Moderate hearing loss | 22.6 % | In both ears | 19.6 % | Pressing your head, neck, or around the ear | 11.5 % |
| Slight hearing loss | 34.9 % | In the head | 12.4 % | Taking a nap | 23.5 % |
| No hearing loss | 21.9 % | More right than left | 20.9 % | Poor sleep quality | 44.8 % |
| **Hyperacusis status** | Ms : 48/194 | Only right | 13.1 % | Driving | 9.8 % |
| Very important problem | 6.2 % | **Rhytmicity *** | Ms : 16/194 | Being stressed or anxious | 50.3 % |
| Important problem | 15.8 % | Follows breathing | 2.2 % | Being relaxed | 0.5 % |
| Moderate problem | 24.7 % | Follows the heart (pulsatile) | 7.3 % | Drinking alcohol | 11.5 % |
| Small problem | 24.0 % | Follows face and jaw movements | 10.1 % | Drinking coffee | 4.9 % |
| No hyperacusis | 29.5 % | Other | 8.4 % | Medications | 6.6 % |
| **Average sleep duration** | Ms : 51/194 | No | 82.0 % | Using hearing aids | 3.3 % |
| Less than 6h | 11.2 % | **Etiology *** | Ms : 16/194 | Other | 6.0 % |
| Around 6h | 32.2 % | Acoustic trauma | 17.4 % | None | 6.6 % |
| Around 7h | 31.5 % | Presbycusis | 19.1 % | **Modulation decrease *** | Ms : 11/194 |
| Around 8h | 21.0 % | Acute otitis | 4.5 % | Very quiet environment | 8.2 % |
| Around 9h or more | 4.2 % | Chronic otitis | 3.4 % | Low intensity sounds | 24.6 % |
| **Initial cause *** | Ms : 16/194 | Serous otitis | 3.4 % | High intensity sounds | 20.2 % |
| Acoustic trauma | 27.0 % | Eustachian tube dysfunction | 1.1 % | Head movements | 1.6 % |
| Hearing loss or hearing trouble | 9.6 % | Congenital hearing loss | 6.2 % | Clenching the teeth or moving the jaw | 2.7 % |
| Ear fullness | 12.4 % | Sudden hearing loss | 5.1 % | Pressing your head, neck, or around the ear | 6.0 % |
| Barotrauma | 4.5 % | Neurinoma | 1.7 % | Taking a nap | 7.1 % |
| Cervical trauma | 3.4 % | Ototoxicity | 3.9 % | Good sleep quality | 29.5 % |
| Concussion | 2.8 % | Ménière disease | 6.2 % | Driving | 11.5 % |
| Stress | 27.0 % | Otosclerosis | 0.0 % | Being stressed or anxious | 1.1 % |
| Illness | 10.7 % | Other | 11.2 % | Being relaxed | 23.0 % |
| No identified cause | 23.6 % | Unknown | 38.8 % | Drinking alcohol | 11.5 % |
| Other | 25.3 % | **Tinnitus variation pattern *** | Ms : 0/194 | Drinking coffee | 1.6 % |
| **Tinnitus pitch** | Ms : 48/194 | From one second to the other (fluctuant) | 16.5 % | Medications | 6.0 % |
| High pitch | 85.6 % | Very different intensities depending days | 30.4 % | Using hearing aids | 20.2 % |
| Medium pitch | 13.0 % | Intense in the morning | 26.8 % | Other | 12.6 % |
| Low pitch | 1.4 % | Intense in the evening | 45.4 % | None | 29.5 % |
| **Tinnitus objective or subjective** | Ms : 49/194 | Other | 8.2 % | | |
| Objective | 9.7 % | No variation | 29.9 % | | |
| Subjective | 90.3 % | | | | |

(*) For these questions, responders could select multiple answers among the proposed choices

*Table 2 – Sample symptomatologic characterization N=194, Abbreviations : VNS : Visual Analog Scale*

### Table 3, Tinnitus, dreams and awakening survey results

| Tinnitus and dreaming | Percentage | Tinnitus and awakening | Percentage |
|---|---|---|---|
| **Do you remember your dreams ?** | Missing : 0/194 | **Tinnitus higher when awakening from a dream** | Missing : 57/160* |
| *I often remember my dreams* | 22.7 % | *No, it decreases* | 1.0 % |
| *I sometimes remember my dreams* | 59.8 % | *No* | 64.1 % |
| *I do not remember my dreams* | 16.0 % | *Yes sometimes* | 21.4 % |
| *I do not dream* | 1.5 % | *Yes, this happens most of the time, I am sure of it* | 13.6 % |
| **Do you hear your tinnitus in your dreams ?** | Missing : 0/160* | **Nocturnal awakening influence on tinnitus** | Missing : 0/194 |
| *I hear my tinnitus in my dreams* | 3.1 % | *Cannot say* | 6.7 % |
| *Sometimes I hear it, sometimes no* | 4.4 % | *Tinnitus appears instantly* | 47.9 % |
| *Not hearing tinnitus in my dreams* | 92.5 % | *My tinnitus wakes me up* | 18.0 % |
| **Dreams about tinnitus ?** | Missing : 0/160* | *Sometimes my tinnitus stays off* | 9.8 % |
| *Yes I, had had a positive dream about my tinnitus* | 2.5 % | *Tinnitus appears gradually* | 14.4 % |
| *Yes I had had a nightmare about my tinnitus* | 12.5 % | *Tinnitus appears abruptly after a while* | 7.2 % |
| *Yes and it often happens* | 4.4 % | *Tinnitus appears when thinking about it* | 15.5 % |
| *No I never had a dream about my tinnitus* | 83.1 % | **Morning awakening influence on tinnitus** | Missing : 0/194 |
| **Do you do lucid dreams ?** | Missing : 0/160* | *Cannot say* | 4.6 % |
| *Yes it has happened to me, but it is rare* | 44.4 % | *Tinnitus appears instantly* | 52.1 % |
| *Yes I often do lucid dreams* | 11.2 % | *My tinnitus wakes me up* | 12.9 % |
| *Yes, and I have trained to have them whenever I want* | 0.6 % | *Sometimes my tinnitus stays off* | 4.6 % |
| *No, I never had a lucid dream* | 43.8 % | *Tinnitus appears gradually* | 21.6 % |
| **Hear tinnitus during lucid dreams ?** | Missing : 111/160* | *Tinnitus appears abruptly after a while* | 4.1 % |
| *Not concerned* | 55.1 % | *Tinnitus appears when thinking about it* | 15.5 % |
| *No, I don't* | 32.7 % | **Nap influence on tinnitus** | Missing : 0/194 |
| *Sometimes I hear it, sometimes no* | 2.0 % | *Cannot say / never do a nap* | 18.6 % |
| *Yes, I do* | 10.2 % | *It decreases my tinnitus* | 13.4 % |
| **Hear external sounds during lucid dreams ?** | Missing : 111/160* | *Doing a nap has no impact on my tinnitus* | 32.5 % |
| *Not concerned* | 61.2 % | *It increases for a moment (10 min, 1 hour) then goes back to normal* | 24.2 % |
| *No, I don't* | 26.5 % | *It increases my tinnitus until the next sleep episode* | 13.9 % |
| *Yes, I do* | 12.2 % | *The sound of my tinnitus changes/a new sound appears* | 6.7 % |
| | | *The same things happen when I extend my sleep in the morning* | 9.8 % |

*(\*) For these questions, only responders that reported remembering at least some of their dreams are reported*

*Table 3 : Tinnitus, Dreams and Awakening survey results*

### Table 4, Cross-table of perceptions during lucid dreams (N = 22)

| | | Hearing tinnitus during lucid dreams | |
|---|---|---|---|
| | | **Yes / sometimes yes, sometimes no** | **No** |
| **Hearing external sounds during lucid dreams** | **Yes** | 7 | 1 |
| | **No** | 1 | 13 |

*Table 4 : Cross-table on participants answers on tinnitus during lucid dreams. Only patients that could report on both questions were selected (patients that reported "not concerned/not being able to report" to either question or both were excluded)*

## Table 5, Symptomatologic characteristics of tinnitus patients hearing tinnitus during dreams

| Symptomatologic characteristic | P-value | Hedge's g | Sample size group 1 | Sample size group 2 |
|---|---|---|---|---|
| Objective tinnitus | p < 0.001 | 1,54 | 8 | 112 |
| ESIT-SQ Initial cause : Serous otitis/Eustachian tube dysfunction | p < 0.001 | 1,34 | 11 | 136 |
| ESIT-SQ hearing pathology : Other | p < 0.001 | 1,23 | 9 | 112 |
| ESIT-SQ : Total deafness / cophosis | p < 0.001 | 1,22 | 9 | 112 |
| ESIT-SQ Initial cause : Illness | p < 0.001 | 1,21 | 11 | 136 |
| ESIT-SQ Initial cause : Aspirin intake | 0,001 | 1,12 | 9 | 112 |
| VNS stress/anxiety (0-10) | p < 0.001 | 1,1 | 12 | 148 |
| ESIT-SQ : Left handed | 0,001 | 1,1 | 9 | 112 |
| ESIT-SQ : Bother duration | 0,024 | 1,1 | 9 | 112 |
| ESIT-SQ : Pulsatile sound following breathing | p < 0.001 | 1,08 | 11 | 136 |
| THI score | 0,001 | 1,03 | 11 | 119 |
| ESIT-SQ hearing pathology : Presbyacousis | 0,003 | 1 | 9 | 112 |
| VNS depressive mood (0-10) | 0,002 | 0,96 | 12 | 148 |
| ESIT-SQ : Using hearing aid and sound generator | 0,005 | 0,93 | 9 | 112 |
| ISI score | 0,003 | 0,92 | 12 | 128 |
| ESIT-SQ hyperacusis : Very important problem | 0,007 | 0,89 | 9 | 112 |
| ESIT-SQ other condition : Hypertension | 0,008 | 0,87 | 9 | 112 |
| ESIT-SQ : Using cochlear implant | 0,011 | 0,84 | 9 | 112 |
| ESIT-SQ Initial cause : Quinin intake | 0,011 | 0,84 | 9 | 112 |
| ESIT-SQ other condition : Other cerebro-vascular issue | 0,011 | 0,84 | 9 | 112 |
| ESIT-SQ : Tinnitus sometimes present (intermittent) | 0,011 | 0,84 | 9 | 112 |
| VNS vertigo (0-10) | 0,007 | 0,79 | 12 | 148 |
| ESIT-SQ : Ear fullness | 0,018 | 0,75 | 9 | 112 |
| VNS headaches (0-10) | 0,012 | 0,75 | 12 | 148 |
| Etiology : Schwanoma | 0,011 | 0,75 | 11 | 136 |
| VNS tinnitus intrusiveness (0-10) | 0,016 | 0,71 | 12 | 148 |
| ESIT-SQ other condition : Sleep onset issue | 0,025 | 0,7 | 9 | 112 |
| ESIT-SQ : Tinnitus progressive emergence | 0,029 | 0,67 | 9 | 112 |
| ESIT-SQ Initial cause : Antibiotic intake | 0,041 | 0,62 | 9 | 112 |
| ESIT-SQ initial cause : Neck trauma | 0,041 | 0,62 | 9 | 112 |
| ESIT-SQ tinnitus presence: Every week | 0,044 | 0,61 | 9 | 112 |
| ESIT-SQ other condition : Diabetis | 0,044 | 0,61 | 9 | 112 |
| ESIT-SQ hearing loss in the family : Son | 0,044 | 0,61 | 9 | 112 |
| ESIT-SQ hearing loss in the family : Brother | 0,043 | 0,61 | 9 | 112 |
| Somatosensory modulation of tinnitus | 0,036 | 0,58 | 11 | 136 |
| ESIT-SQ modulation : Decrease when jaw clenching | 0,042 | 0,56 | 11 | 139 |
| Tinnitus sound : Melody | 0,046 | 0,54 | 11 | 134 |
| ESIT-SQ Initial cause : No drug intake | 0,006 | -0,91 | 9 | 112 |

*VNS : Visual Numeric Scale, THI : Tinnitus Handicap Inventory, ISI : Insomnia Severity Index, ESIT-SQ : European School of Interdisciplinary Tinnitus research - Screening Questionnaire*

*Table 5 : Symptomatologic characteristics of tinnitus patients hearing tinnitus during dreams. Group 1 : Patients reporting hearing tinnitus during dreams, Group 2 : patients reporting not reporting hearing tinnitus during dreams. Negative Hedge's g means a characteristic more present in group 2 than group 1. VNS : Visual Numeric Scale, THI : Tinnitus Handicap Inventory, ISI : Insomnia Severity Index, ESIT-SQ : European School of Interdisciplinary Tinnitus research - Screening Questionnaire*

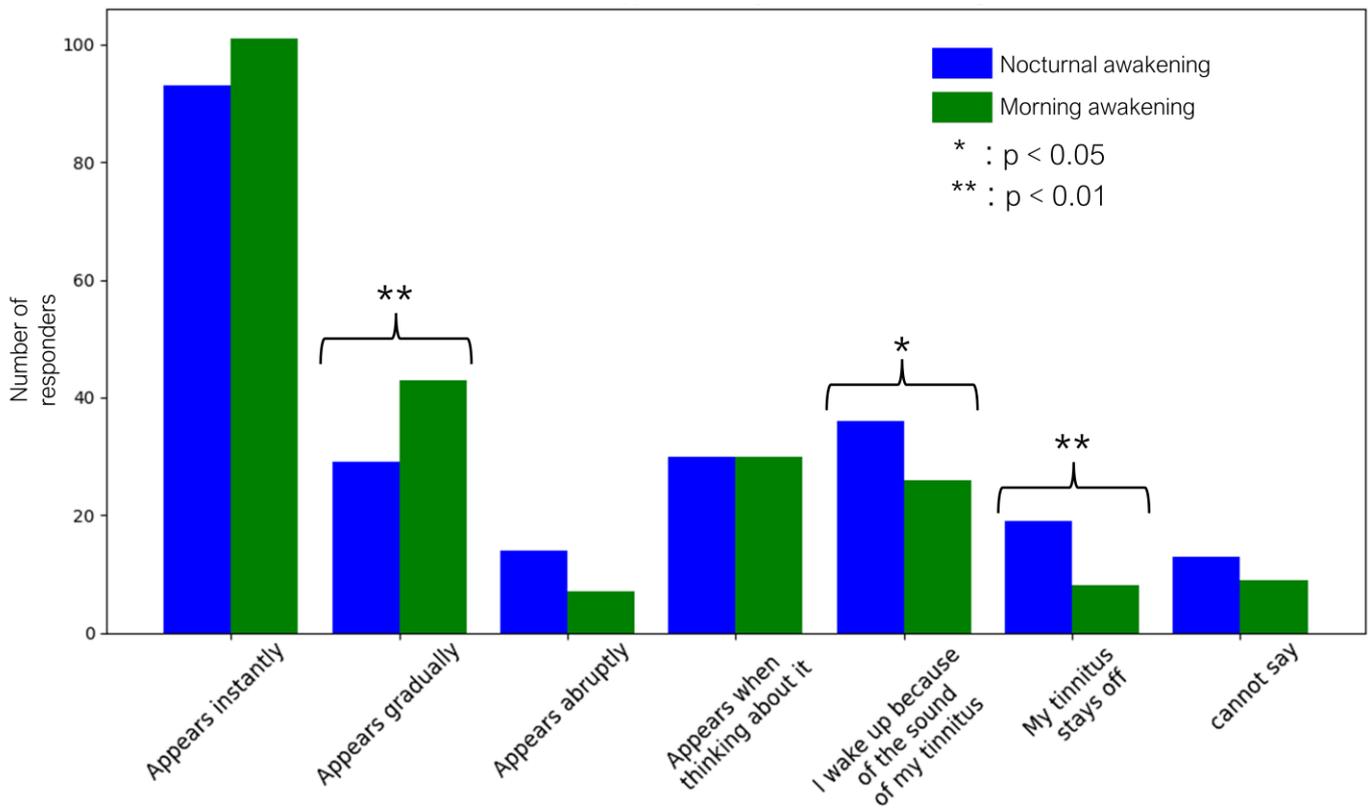

*Figure 1: Survey results on tinnitus temporality of manifestation upon awakening during nocturnal awakenings and in the morning (N=194).*

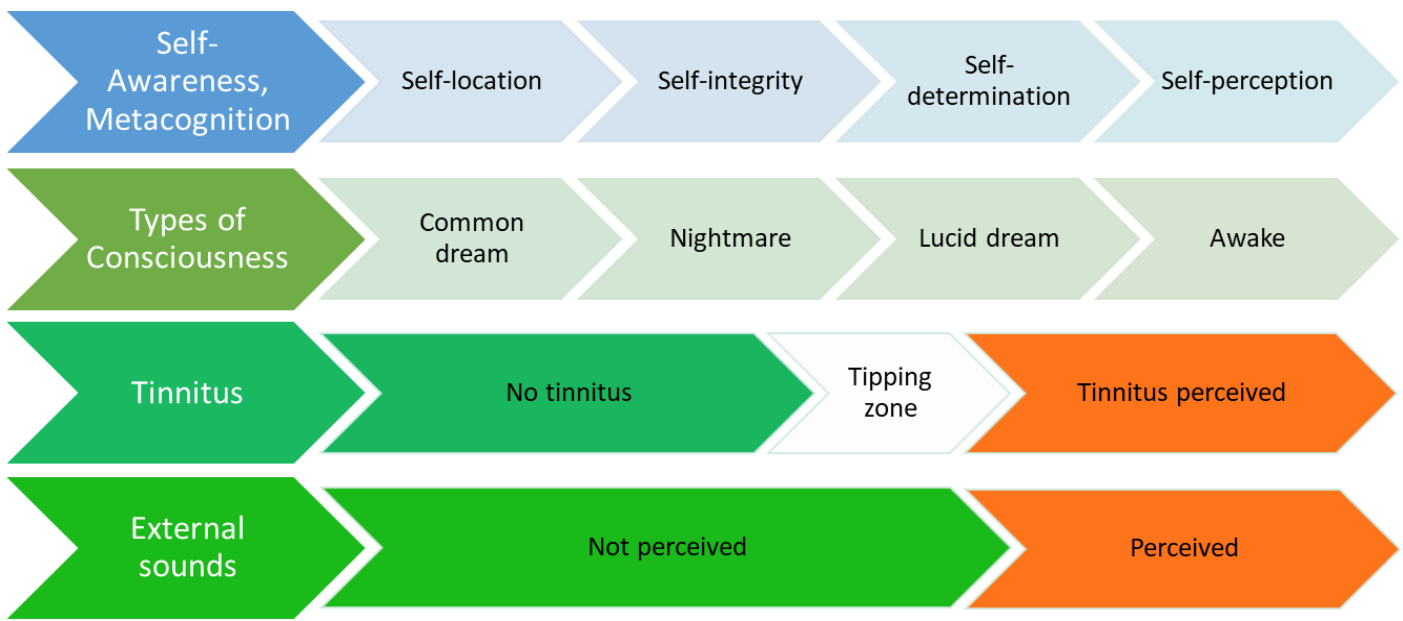

*Figure 2 : A theoretical account of the heterogeneity of tinnitus experience in dreaming and waking consciousness according to the complexity of self-awareness. The first axis presents 4 levels of self-awareness by order of complexity, from simple self-location in common dreams (i.e., ''a purely spatiotemporal first-person perspective'' according to* (Windt, 2010)) *to sensory-based awareness of self-perception* (Candia-Rivera, 2024). *Between these two levels, self-integrity specifies awareness of oneself in nightmares experienced as threat to self-boundaries, and self-determination specifies the dreamer's sense of agency in lucid dreams* (Dresler et al., 2014). *The second axis distinguishes 4 types of consciousness that are discussed in the present study: common dreams, nightmares, lucid dreams, and wakefulness. Contrasted experience of tinnitus percept in wakefulness and its absence in common dreams and most nightmares are related to sensory-inputs in self-perception vs sensory-isolation in dreaming consciousness. The encounter of hindrances in the dream plot (i.e., feedback) that can occur lucid dreams and maybe in some nightmares is suggested as a tipping zone for the reemergence of tinnitus, which becomes again an 'attribute' of self-awareness. This heightened self-awareness would be accompanied with a failure of gating external auditory stimuli, that is associated in participants' reports with the presence of tinnitus in self-perception.*

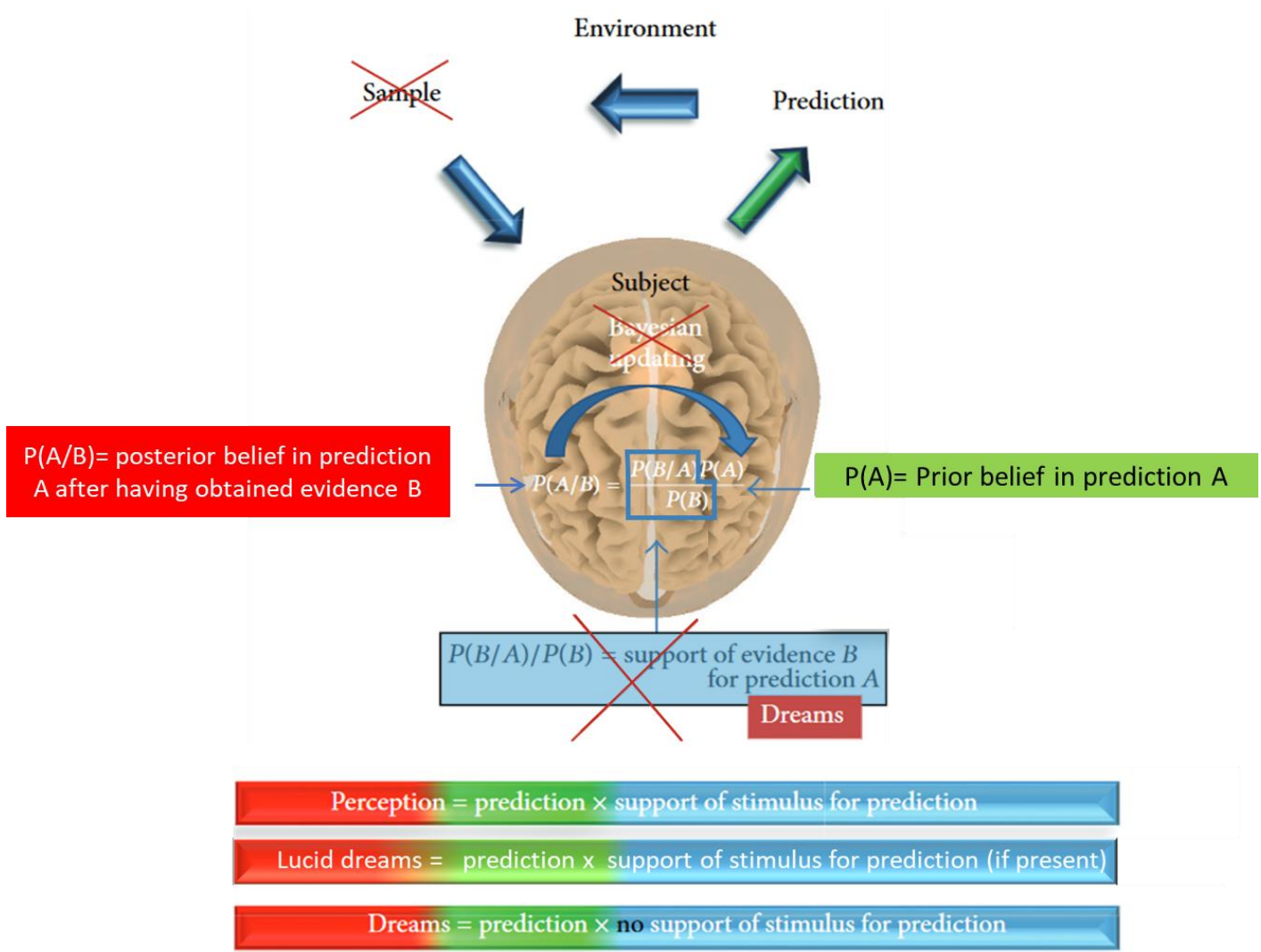

*Figure 3 : A revised version explanation of the absence of tinnitus in dreams according to the Bayesian brain model for tinnitus, as presented in* (De Ridder et al., 2014a), *including the novel findings concerning lucid dreaming.*

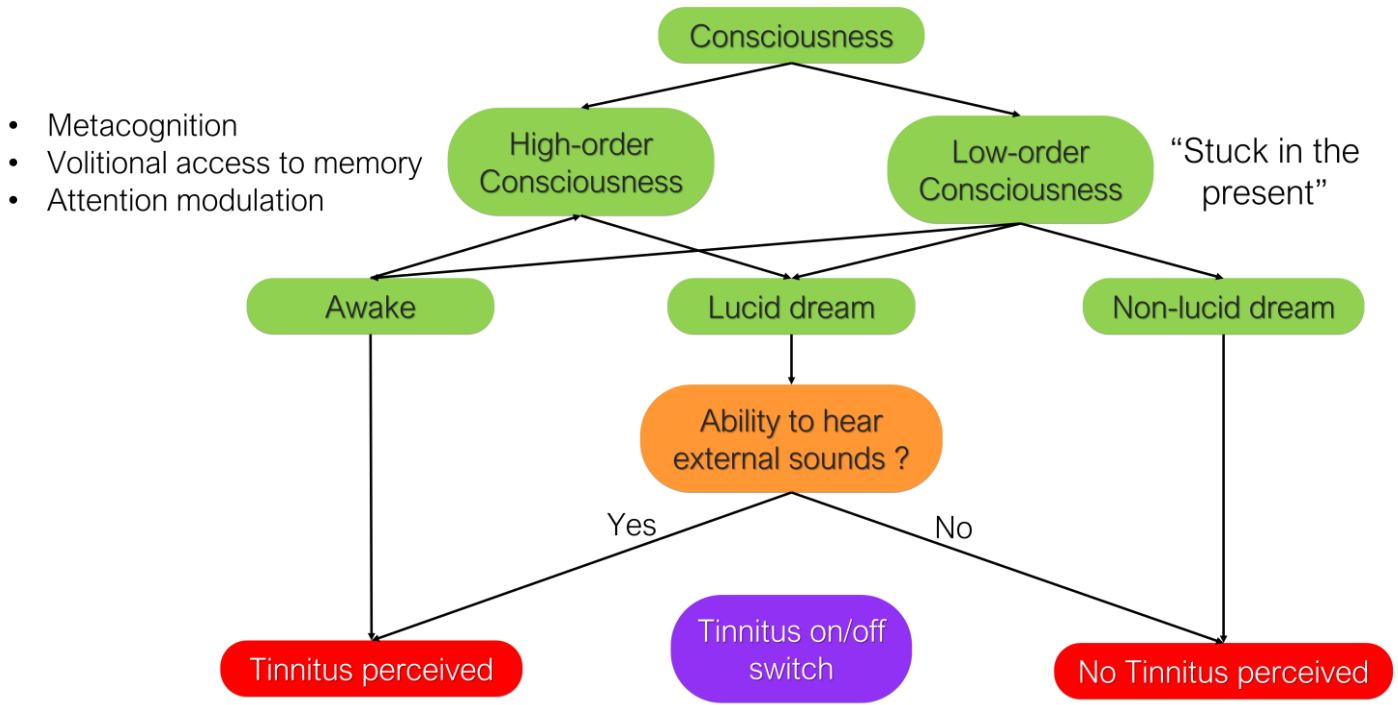

*Figure 4: Revisited proposed model for characterization of tinnitus on/off switch location according to the integrative model of auditory phantom perception* (De Ridder et al., 2014b)*, taking into account observations made on lucid dreamers tinnitus perception.*

*Supplementary Material:*

# Qualitative analysis of interviews with lucid dreamer tinnitus patients

This preliminary exploration of tinnitus in relation to lucid dreaming was carried out with 5 participants who agreed to share their experience in semi-directive interview with a clinical psychologist (ND). These interviews lasted between 30 and 65 minutes, depending on the participants, and were recorded to enable the reports below.

Four themes were systematically investigated with each participant:

1° their experience of tinnitus during the day (i.e. if they can forget about it, or not hear it, and how they deal with intrusion) ; 2° their experience of tinnitus during the night (as an auditory percept and a subject of concern) in lucid and common dreams; 3° their degree of lucidity in their dreams (i.e. awareness, self-reflection , and sense of agency) and its potential influence on their experience of tinnitus; 4° their perceptions outside the dream frame (e.g. body position, external sounds) and their potential influence on their experience of tinnitus.

These themes were explored with the aim of clarifying the individual experience of perceiving (or not perceiving) tinnitus during participants' recall of their dreams.

The main findings of this preliminary investigation are:

1. Lucid dreamers do not report hearing tinnitus or worrying about having tinnitus when they have common dreams (i.e. non-lucid dreams).

2. Access to metacognition in lucid dreaming (awareness, self-reflection, sense of agency) does not lead to the perception of tinnitus, provided that there is no interruption in the individual's motivation to see the dream through.

3. When lucid dreamers find themselves unable to pursue compelling goals (e.g. flying or escaping a threat), the failure of their will is accompanied by the return of tinnitus into their lucid dream and is followed by awakening.

4. When lucid dreaming is interrupted by awareness of stimuli outside the dream frame, such as real sounds in the environment, perception of tinnitus returns to consciousness and is followed by awakening. Awareness of real external stimuli that do not interrupt dreaming can occur without tinnitus being present in the lucid dream.

### Case 1. Female participant, 49 year-old (tinnitus duration : 10 years)

She became interested in lucid dreaming as a young adult. When a lucid dream happens to her, she always wants the dream to continue, to watch it unfold with great interest. But she cannot change the course of the dream or even carry out certain actions voluntarily. She is particularly aware of how fragile her lucidity is during a dream, and how difficult it is to 'stay in the dream' and not wake up quickly. She remembers very little about the content of her dreams.

One day, she realized that she could not hear her tinnitus in her dreams. She describes this situation as 'two separate worlds', like someone who loses their sight but continues to see in their dreams as if nothing had happened. In fact, she has no recollection of ever worrying that she 'had tinnitus' while she was dreaming. This fact presents itself to her without any effort or reflection on her part. However, she finds that her sleep is not restorative and that she is more tired since the onset of her tinnitus. During the day, she feels a 'presence' that is sometimes less obvious, but 'always there'.

She once began a lucid dream, early in the night, and remembers hearing the sound of a sliding door being pushed open by her husband. She concentrated on not paying attention to what she perceived outside her dream, and managed not to wake up. She does not remember hearing her tinnitus while she was still dreaming. On another occasion, as her husband was getting up early in the morning, she tried in vain to convince herself not to 'get out' of her dream. She remembers hearing her tinnitus in her dream at that moment, before waking up for good. The situation is very different in non-lucid dreams, as she 'does not force' her attention on the course of the dream or even realize that what she is seeing is not real.

**Case 2. Male participant, 26 year-old (tinnitus duration : 4 years)**

His first memory of a lucid dream dates back to his childhood. He became very interested in lucid dreaming as a young adult, after learning that he could change the course of his dreams. He discovered that he should not change it too much, or oppose what is happening in a dream, otherwise he runs the risk of waking up. There is a 'balance' to be respected and he should not experiment too much, otherwise he becomes aware of 'what is really going on around you'. He remembers his dreams in detail, such as one time when he was in a war scene and was shot. Not being dead, he realized he was dreaming and transformed 'into the Hulk' to destroy his opponents with great satisfaction.

He is certain that he never heard his tinnitus during his dreams. He has no recollection of thinking about it either, as he soon realized that there was no medical solution and did not seek advice or help. Soon after the onset of his tinnitus, he was happy to go to sleep because he knew he would not hear it, and so could 'take a break'. When he is lucid and acting on a dream, he does not hear his tinnitus. It is only after he has changed too many things that he wakes up and, at that moment, hears his tinnitus again. He remembers realizing he was dreaming because he could not hear his tinnitus. As with the war scene (above), it was the incongruence with reality that made him realize he was actually dreaming.

Tinnitus is no longer a problem for him today. He can forget about it effortlessly, and no longer worries about hearing it at the end of the day, in the quiet. He convinced himself that the presence of his tinnitus would help him fall asleep, as if it were a friend at his side. It is only when he is tired that his tinnitus becomes more annoying, but without causing any particular concern.

**Case 3. Male participant, 55 year-old (tinnitus duration : 2 years)**

He has been having recurrent nightmares for years, in which he is frightened by an unknown presence that is stronger than he is. He hears himself screaming in his nightmares. This usually wakes him up, as well as his relatives. He sometimes has 'tinnitus nightmares' in which he hears his tinnitus while he is asleep, just as when he is awake. Tinnitus then gets louder and louder until he wakes up screaming. In contrast, when he is having lucid dreams without being in danger, he does not perceive his tinnitus. If he takes sufficient care not to fly too high (for fear of heights), he can 'take long trips' as if flying in an airplane, without his tinnitus being present.

He had no explanation for the sudden onset of his tinnitus. The only concomitance he could identify was his covid vaccination 10 days before, but he had no certainty about this. Following a prescription for sleeping pills, he noticed that it was easier for him to focus throughout the day. A polysomnography revealed poor quality of sleep, with more than 30 micro-awakenings per hour. Sleep management with medication is essential in his daily life with tinnitus.

He recalled a recent nightmare in which he heard his tinnitus, with a surprising transition to lucid dreaming. First he was pursued by an unknown presence from which he wanted to escape by flying. As they tried to hold him back by his clothes, he had the sensation of his own body lying in bed ('the physical sensation of myself') and then decided lucidly to climb stairs to escape the danger. As he did so voluntarily, he remembers hearing his tinnitus. He thinks he had already heard it in the first part of the nightmare, but as an 'external noise' comparable to an alarm. It was only when he climbed the stairs that he identified it as coming from his own body, before waking up hearing his tinnitus.

**Case 4. Male participant, 49 year-old (tinnitus duration : 3 years)**

For one year after the onset of his tinnitus, he was afraid to go to sleep during the day, because when he woke up tinnitus loudness was 'double'. This was the case even after a few minutes nap. He fell asleep listening to music or white noises. At night when he woke up, he immediately perceived his tinnitus intensely. Since three years now, he 'does not know what silence is anymore'. He can't remember his dreams well, but he does not believe having ever heard his tinnitus when he was asleep. Conversely, he is certain that he has never had a dream 'about not having tinnitus'.

After a year, the 'rebound effect' in tinnitus loudness had disappeared. Loudness is now constant before and after falling asleep. He has no explanation for this phenomenon. He does not 'hear it' anymore during a conversation, before tinnitus suddenly 'appears' to his consciousness when he becomes aware of it. When he wakes up, he also has the impression of a clear transition: 'from nothing to something' (i.e. from silence to tinnitus). He feels frustrated at not having an explanation for the onset of tinnitus and its interaction with his sleep.

One morning he had a very soothing experience, which he later relived. In a state between sleep and wake, he noticed that his tinnitus 'was not there'. He is not awake at these times, but he is aware of sensations such as the warmth of the bed or the weight of his lying body. He can also 'hear something' around him, but remains indifferent to what is going on, as if he was 'cut off from the world' and 'out of time'. This happens when he can stay in bed to make graceful morning, without any professional constraints. Nothing frightening or fantastic happens in this state of consciousness. He dreams mainly of 'everyday thoughts' that are associated with a 'state of well-being'. He welcomes this experience of being free from tinnitus, without trying to understand it, letting himself be 'carried away by his daydreams'.

**Case 5. Female participant, 77 year-old (tinnitus duration : 40 years)**

She loves flying in her lucid dreams because it gives her a feeling of freedom. When she can't, she feels deprived of a joy she would like to prolong as long as possible. While she is having common (non-lucid) dreams, she can't hear her tinnitus. However, she easily 'switches' into lucid dreaming, when she remarks to herself about what she sees (e.g. 'the dream isn't very beautiful, we should move on'). The absence of tinnitus can continue into lucid dreaming when she is amazed by what she is exploring. For example, she has found herself in the 1950s, with her (deceased) father at her side as if she was a child, finding it strange not to see cars in the streets. The absence of tinnitus at these times is clear to her.

Tinnitus is a 'presence' that has been with her for 45 years. She has learned to put it aside when she is not having a crisis. She suffers from wide variations in intensity, aggravated by fatigue and alcohol. Her tinnitus can disturb her ability to listen to conversations or enjoy birdsong. Until a few years ago, she had 'islands of silence' that she found amazing, but which no longer exist. She would like to finally be free of tinnitus, which she still suffers greatly from some days.

While being pursued by 'enemies' in nightmares, she was able to escape them by voluntarily flying high into the clouds. When she flies, however, her experience can be highly contradictory. Climbing higher and higher so that there is 'nothing around her', tinnitus returns in her dreams, also getting louder and louder, to the point where she suddenly wakes up. She then finds herself with tinnitus and very bothered by her physical environment. Therefore, she 'must not overdo it' by trying to fly too high, as her tinnitus 'always gets the better of her'.